\documentclass{midl}

\pdfoutput=1
\usepackage[nolist]{acronym}
\usepackage{chngcntr} % used to label appendix figures
\usepackage[percent]{overpic} % used to overlay text on figures
\usepackage{float}
\usepackage{capt-of}

\jmlryear{2022}
\jmlrworkshop{Full Paper -- MIDL 2022}

\title[DMRI Angular Super-Resolution with a 3D RCNN]{
    Angular Super-Resolution in Diffusion MRI with a 3D Recurrent Convolutional Autoencoder
}

\midlauthor{\Name{Matthew Lyon\nametag{$^{1}$}} \Email{matthew.lyon-3@postgrad.manchester.ac.uk}\\
\Name{Paul Armitage\nametag{$^{2}$}} \Email{p.armitage@sheffield.ac.uk}\\
\Name{Mauricio A Álvarez\nametag{$^{1}$}} \Email{mauricio.alvarezlopez@manchester.ac.uk}\\
\addr $^{1}$ Dept. of Computer Science, University of Manchester, UK\\
\addr $^{2}$ Dept. of Infection, Immunity and Cardiovascular Disease, University of Sheffield, UK\\
}

\begin{document}

\maketitle

\begin{abstract}
    High resolution diffusion MRI (dMRI) data is often constrained by limited scanning time
    in clinical settings, thus restricting the use of downstream analysis techniques that
    would otherwise be available. In this work we develop a 3D recurrent convolutional neural
    network (RCNN) capable of super-resolving dMRI volumes in the angular (q-space)
    domain. Our approach formulates the task of angular super-resolution as a patch-wise regression
    using a 3D autoencoder conditioned on target b-vectors. Within the network we use a
    convolutional long short term memory (ConvLSTM) cell to model the relationship between
    q-space samples. We compare model performance against a baseline spherical harmonic
    interpolation and a 1D variant of the model architecture. We show that the 3D model has the
    lowest error rates across different subsampling schemes and b-values. The relative performance
    of the 3D RCNN is greatest in the very low angular resolution domain. Code for this project is
    available at \href{https://github.com/m-lyon/dMRI-RCNN}{github.com/m-lyon/dMRI-RCNN}.
\end{abstract}


\begin{acronym}
    \acro{ae}[AE]{absolute error}
    \acro{clstm}[ConvLSTM]{convolutional long short term memory}
    \acro{cnn}[CNN]{convolutional neural network}
    \acro{dmri}[dMRI]{diffusion MRI}
    \acro{dti}[DTI]{diffusion tensor imaging}
    \acro{fa}[FA]{fractional anisotropy}
    \acro{fod}[FOD]{fibre orientation distribution}
    \acro{gm}[GM]{grey matter}
    \acro{hardi}[HARDI]{high angular resolution diffusion imaging}
    \acro{hcp}[HCP]{WU-Minn Human Connectome Project}
    \acro{lar}[LAR]{low angular resolution}
    \acro{lstm}[LSTM]{long short term memory}
    \acro{mae}[MAE]{mean absolute error}
    \acro{mri}[MRI]{magnetic resonance imaging}
    \acro{noddi}[NODDI]{neurite orientation dispersion and density imaging}
    \acro{rcnn}[RCNN]{recurrent \ac{cnn}}
    \acro{rmse}[RMSE]{root mean squared error}
    \acro{psnr}[PSNR]{peak signal-to-noise ratio}
    \acro{mssim}[MSSIM]{mean structural similarity index measure}
    \acro{ssr}[SSR]{spatial super-resolution}
    \acro{sar}[ASR]{angular super-resolution}
    \acro{sh}[SH]{spherical harmonics}
    \acro{sr}[SR]{super-resolution}
    \acro{wm}[WM]{white matter}
    \acro{relu}[ReLU]{Rectified Linear Unit}
\end{acronym}

\section{Introduction}

Advances in \ac{dmri} analysis techniques continue to push the boundaries of what
is attainable through the non-invasive imaging modality
\citep{zhang2012noddi,raffelt2017investigating,drake2018clinical}. However, acquiring the
\ac{hardi} that is needed for these more advanced techniques presents a challenge. \ac{hardi} data
requires the acquisition of typically thirty or more diffusion directions, often at several
b-values (multi-shell), to use these techniques effectively. It is therefore clinically infeasible
to benefit from these advances due to the time constraints of acquiring such high resolution
datasets.

One way to reduce the burden of acquisition time is through the use of image
enhancement techniques such as \ac{sr}. Here \ac{dmri} data has two distinct, but related,
resolutions that can be super resolved: spatial resolution, or the density of sampling within
k-space, and angular resolution, or the density of sampling within q-space \citep{tuch2004q}.
\Ac{ssr} has been extensively covered in the medical imaging and natural image domains
\citep{li2021review,yang2019deep}. However, as \ac{dmri} data has a unique angular
structure, the amount of work done in the \ac{sar} domain is relatively limited.

In particular, many methodologies opt to constrain the challenges of \ac{sar} by performing
inference on downstream analysis techniques. This has the advantage of simplifying the task, but
limits the ability of the super-resolved data to be used in different analysis techniques.
Both \citet{lucena2020enhancing} and \citet{zeng2021fod} used single-shell data and \ac{cnn}
architectures to infer \ac{fod} data with similar quality to a multi-shell acquisition
\citep{tournier2007robust}. Similarly, \citet{golkov2016q}, \citet{chen2020estimating}, and
\citet{ye2020improved} developed deep architectures to infer metrics from models such as \ac{noddi}
\citep{zhang2012noddi} and others, that would otherwise be unavailable with single-shell data.

Models that work with diffusion data directly often do so with the use of \ac{sh}
\citep{frank2002characterization}. \ac{sh} provide a set of smooth basis functions defined on the
surface of a sphere. As they form a complete orthonormal basis, they can be used to describe
any well behaving spherical function. As such, they are commonly used to represent \ac{dmri}
signal, which is measured at different diffusion directions defined by points (b-vectors) on the
surface of a unit sphere. Typically within \ac{dmri} deep learning the \ac{sh} coefficients are
first fit to the diffusion data, and then used as input in place of the unconstrained diffusion
signal. The network can then be trained to infer the \ac{sh} coefficients of other shells, as the
\ac{sh} framework already provides interpolation to other data points within a single shell. For
example, \citet{koppers2016diffusion} used \ac{sh} coefficients from single-shell \ac{dmri} data to
infer \ac{sh} coefficients of a different shell. This method was limited in scope however,
as only randomly sampled \ac{wm} voxels within the brain were used. \citet{jha2020multi} then
extended this idea using a 2D \ac{cnn} autoencoder architecture, that inferred data across the
whole brain. Currently only one other deep learning architecture proposed by \citet{yin2019fast}
infers raw \ac{dmri} data without the use of \ac{sh}. This architecture is a 1D \ac{cnn}
autoencoder, which therefore does not benefit from the spatial relationships present within
\ac{dmri} data.

This paper proposes a novel implementation of \ac{sar} in \ac{dmri} data through the use of a
\ac{rcnn} autoencoder architecture. This involves two key innovations: 1) extending the
dimensionality of the network to 3D; 2) using a 3D \ac{clstm} cell to model the q-space
relationships. Both of these contributions allow us to leverage the spatial correlations present
within \ac{dmri} data to efficiently infer new diffusion directions without the
constraint of predefined functions such as \ac{sh}. Additionally, omitting the \ac{sh} framework
allows us to explore the feasibility of using unconstrained \ac{dmri} data in deep learning
inference.

We evaluate the performance of the proposed model by measuring the deviation of \ac{dmri}
signal from the ground truth across multiple diffusion directions. The \ac{hcp} dataset
\citep{van2013wu} is used for training and quantitative comparison. We evaluate model
performance across different angular resolutions and b-values. Additionally, we compare
results from our proposed 3D model with angular interpolation within the \ac{sh} framework,
and a 1D variant of the same model.

\section{Methods}

We formulate the task of \ac{sar} in the following way: \ac{lar} datasets, comprising of 3D
\ac{dmri} volumes and b-vectors, are used as context data to generate a latent
representation of the entire q-space. This latent representation is then queried with target
b-vectors, to infer previously unseen \ac{dmri} volumes. We list below the pre-processing steps
required and network implementation.

\subsection{Pre-processing}
The \ac{hcp} \ac{dmri} data is used for both training and evaluation, and is initially
processed with the standard \ac{hcp} pre-processing pipeline \citep{glasser2013minimal}.
Each 4D \ac{dmri} volume within each subject in the \ac{hcp} dataset contains three shells of
b-values 1000, 2000, and 3000. Each shell is processed independently and contains 90 diffusion
directions, of which the \ac{lar} dataset are subsampled from. Several further pre-processing
steps are necessary to transform the data into an appropriate format for efficient training within
the network. First \ac{dmri} data are denoised. Noise is assumed to be independent across the
entire 4D volume, therefore noise within the context dataset cannot be used to predict the noise
within the target volumes. To mitigate this problem, a full-rank locally linear denoising
algorithm known as `patch2self' \citep{fadnavis2020patch2self} is applied to the data. Next,
\ac{dmri} data are rescaled such that the majority of the distribution lies between $[0,1]$.
This is done by dividing each voxel by a normalisation value given it's shell membership.
$4000$, $3000$, and $2000$ were found to work well as normalisation values for the shells
$b = 1000$, $b = 2000$, and $b = 3000$ respectively.

Afterwards, \ac{dmri} data are split into smaller patches with spatial dimensions
$(10\times10\times10)$. This is done to mitigate the memory limitations of using 4D data, which
would have a prohibitively large memory requirement if kept at full size. $10^{3}$ isotropic was
found to be a large enough patch size to benefit from non pointwise convolutions, whilst still
having reasonable memory requirements. Each patch contained at least one voxel from a brain
extracted mask, thus patches which contained no voxels within the brain are discarded.

Next, during training only, the q-space dimension within \ac{dmri} patches and b-vectors are
shuffled. This is a crucial step in encouraging the model to learn the relationship between the
measured \ac{dmri} signal and the b-vector directionality, whilst additionally discouraging the
model from converging on a solution that is sensitive to the order of the q-space samples. As
such, the shuffling process is repeated after each training epoch. As the training examples are
subsampled from the full 90 directions, it is important to ensure that the q-space shuffling
produces examples that are approximately evenly distributed across the q-space sphere. To do this,
an initial direction is chosen at random, then, the next points are sequentially chosen to minimise
the total angular distance between all points previously selected. This is repeated until the
number of points equals the training example size. Finally, the shuffled datasets are split into
context and target sets of size $q_{in}$ and $q_{out}$ respectively.

\subsection{Proposed Network}

\begin{figure}[!htb]
    \includegraphics[width=\textwidth]{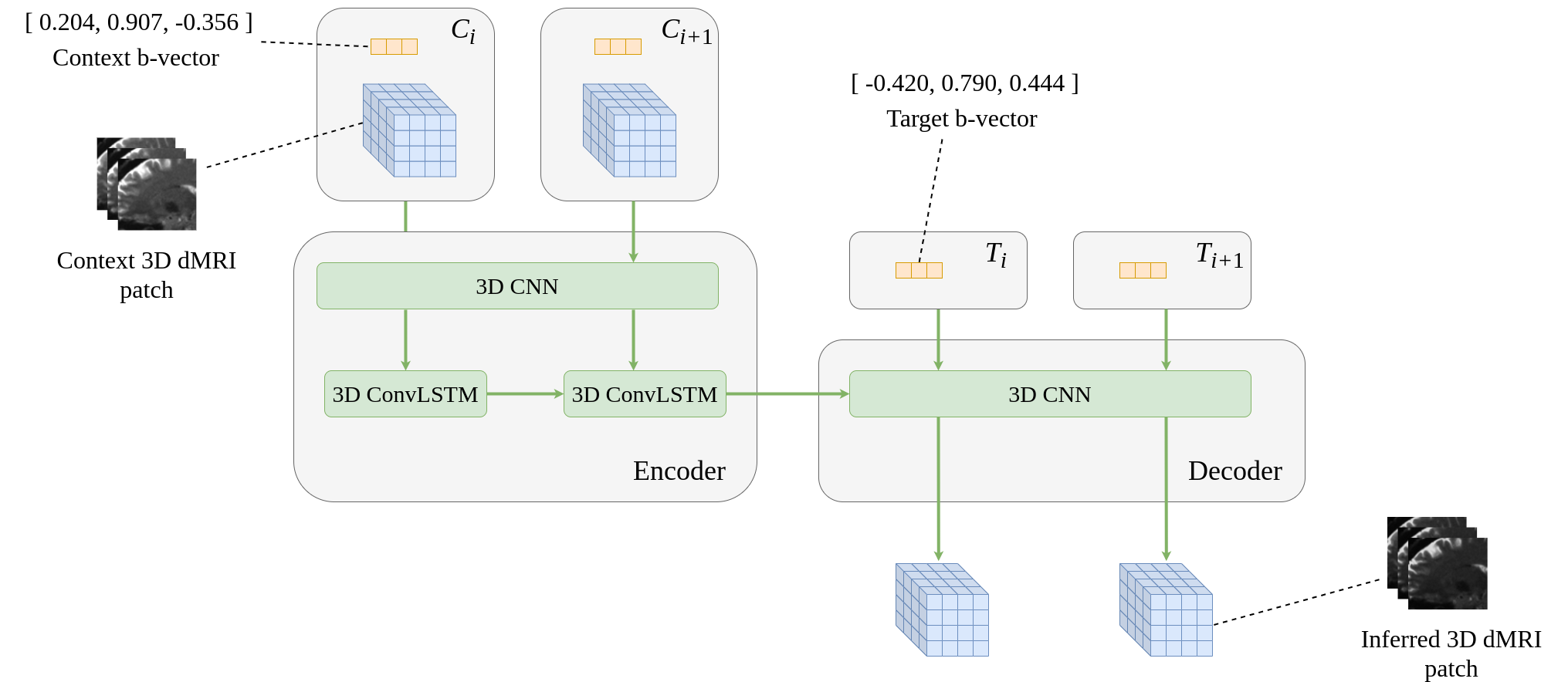}
    \caption{
        \ac{rcnn} model design. Here q-space context data $C_{i}$ are given to the encoder
        sequentially until all context examples $C_{q_{in}}$ are seen. Next, the internal hidden
        state of the \ac{clstm} is passed to the 3D CNN decoder along with target data $T_{i}$
        to infer 3D \ac{dmri} patches along the given diffusion direction.
    }
    \label{fig:rcnn_model}
\end{figure}

The proposed \ac{rcnn} is a conditional autoencoder, comprising of a 3D \ac{cnn} encoder and
decoder. A graphical representation of the architecture is shown in Figure \ref{fig:rcnn_model}.

\subsubsection*{Encoder}

Each encoder input consists of a context set, of size $q_{in}$, containing \ac{dmri}
patches and corresponding b-vectors. Initially within the encoder, the set of b-vectors are
repeated in each spatial dimension, then concatenated with the \ac{dmri} signal, channel-wise,
to form a `q-space tensor'. Next, the q-space tensor is passed through a pointwise
convolutional layer, then onto two parallel convolution blocks connected in series. Afterwards,
the output is passed through two more pointwise convolutional layers, and finally onto a 3D
\ac{clstm} layer. As the convolutional layers within the encoder contain both \ac{dmri} signal
and b-vector information, this allows the encoder network to directly learn the q-space
representation of the \ac{dmri} data.

\subsubsection*{Decoder}

The hidden internal state within the \ac{clstm} layer, alongside a set of target b-vectors, is used
as input to the decoder. First, the hidden state is repeated $q_{out}$ times, where $q_{out}$ is
the number of target b-vectors. Then, similarly to the encoder, the target b-vectors are repeated
in each spatial dimension and concatenated with the hidden state channel-wise. Afterwards the
resultant tensor is passed through two pointwise convolutional layers, then subsequently
concatenated with the target b-vectors again. Finally this is passed through two
parallel convolutional blocks, and subsequently two convolutional layers.

\subsubsection*{Parallel Convolution Blocks}
Parallel convolution blocks are used within the encoder and decoder. They consist of three
convolution layers with kernel sizes $(1\times1\times1)$, $(2\times2\times2)$, and
$(3\times3\times3)$, that apply a convolution operation to the same input in parallel. The
$2^{3}$ and $3^{3}$ isotropic kernels have padding applied prior to the convolution operation, such
that the resultant shape is equal to the unpadded input tensor. This block is inspired by work
done in \citet{szegedy2016rethinking}, and allows for different resolutions of spatial information
to pass through the block in tandem. Outputs from the three convolutions are then concatenated
together channel-wise with an additional residual input. Within the encoder, this residual input
is the aforementioned q-space tensor, and within the decoder it is the target b-vectors that have
been repeated across spatial dimensions. By including this additional input, b-vector
directionality and image information has straightforward propagation throughout the network,
allowing the model to more easily learn the complex q-space relationship within the data.

\subsubsection*{Implementation Details}
All convolutional layers, excluding the \ac{clstm} layer and the final two convolutional layers
within the decoder, consist of the following: a convolution operation, a Swish activation function
\citep{ramachandran2017searching}, and either instance or batch normalisation. The final two
decoder layers have no normalisation and, as \ac{dmri} is strictly positive, \ac{relu}
activation is used in the final layer in place of a Swish activation. The \ac{clstm} is a 3D
extension of the 2D \ac{clstm} presented in \citet{shi2015convolutional}, where the dense
connections of a standard \ac{lstm} cell are replaced with convolutional kernels. It uses standard
activations found within a \ac{lstm} cell and no normalisation. Each of the convolutional layers,
except for the \ac{clstm} layer, treat the q-space dimension as an additional batch dimension,
therefore sharing parameters across q-space samples. All convolutional layers use a stride of
$(1\times1\times1)$. Hyperparameters used for each layer can be found in Figure
\ref{fig:rcnn_full}, and were obtained by a hyperparameter search using KerasTuner
\citep{omalley2019kerastuner} and the Hyperband algorithm \citep{li2017hyperband}. Models were
trained for 120 epochs using the optimizer Adam \citep{kingma2014adam} with \ac{mae} loss function
and a learning rate of $0.001$. The weights used for analysis were the best performing within the
validation dataset. Training and validation datasets comprised of data from 27 and 3 \ac{hcp}
subjects, respectively.

\subsection{SH Q-Space Interpolation}

To interpolate \ac{dmri} data using \ac{sh}, first \ac{sh} coefficients $\mathbf{c}_{sh}$
for each spatial voxel are found using the pseudo-inverse least squares method in equation
\eqref{eq:sinv} below,

\begin{equation} \label{eq:sinv}
    \mathbf{c}_{sh} = (\mathbf{B}_{L}^\top \mathbf{B}_{L})^{-1} \mathbf{B}_{L}^\top \mathbf{s}_{L}.
\end{equation}

Here, $\mathbf{B}_{L}$ is a matrix that denotes the \ac{sh} basis for the low angular resolution
dataset, where each row contains the \ac{sh} expansion sampled at a given diffusion direction.
$\mathbf{s}_{L}$ is a vector containing the measured signal voxel at various diffusion
directions. The full resolution dataset $\mathbf{s}_{H}$ is reconstructed simply via equation
\eqref{eq:recon}, where $\mathbf{B}_{H}$ is the \ac{sh} basis matrix containing all
diffusion directions within the shell,

\begin{equation} \label{eq:recon}
    \mathbf{s}_{H} = \mathbf{B}_{H} \mathbf{c}_{sh}.
\end{equation}

The set of \ac{sh} basis functions sampled to obtain $\mathbf{B}_{L}$ and $\mathbf{B}_{H}$ are the
modified \ac{sh} functions $\widetilde{Y}_{l}^{m}(\theta,\phi)$ first defined in
\citet{tournier2007robust}:

\begin{equation}
    \widetilde{Y}_{l}^{m}(\theta,\phi) = \begin{cases}
        0                                       & \text{if $l$ is odd}, \\
        \sqrt{2} \: \Im (Y_l^{-m}(\theta,\phi)) & \text{if $m < 0$},    \\
        Y_l^0(\theta,\phi)                      & \text{if $m = 0$},    \\
        \sqrt{2} \: \Re (Y_l^m(\theta,\phi))    & \text{if $m > 0$},    \\
    \end{cases}
\end{equation}

where $Y_{l}^{m}(\theta,\phi)$ defines the \ac{sh} basis function of order $l$ and $m$.

\section{Experiments and Results}

We evaluate the performance of our model on eight previously unseen subjects from the \ac{hcp}
dataset for three diffusion shells and at varying q-space undersampling ratios. Each result in
the presented tables is obtained from a separately trained model, with the same architecture and
hyperparameters, except for the 3D \ac{rcnn} (Combined) model. Instead, this model is trained with
all three b-values concurrently. As a baseline comparison, results from the \ac{rcnn} models are
compared to \ac{sh} interpolation. Here, a maximum \ac{sh} order of 2 is used, as this was found
to produce the most accurate reconstruction within the subsampling ratios used. \Ac{rmse}, and
\ac{mssim} results are presented with respect to the ground truth of the measured diffusion
directions for the eight subjects. \ac{rmse} and \ac{mssim} in the presented tables are given as a
mean and standard deviation across q-space samples in the eight subjects. Each value within these
error distributions is averaged across all spatial dimensions within each q-space sample and
subject.

Table \ref{table:subsamp} compares performance at different subsampling ratios across the three
\ac{sar} models. The proposed 3D \ac{rcnn} outperforms both the 1D variant and \ac{sh}
interpolation across all three subsampling ratios in \ac{rmse} and \ac{mssim}. The largest
relative gain in performance is present at the lowest subsampling ratio $q_{in} = 6$. Here the
3D \ac{rcnn} has a reduction in \ac{rmse} of 34.1\% compared to \ac{sh} interpolation,
and a reduction in the \ac{rmse} standard deviation of 72.4\%. This suggests that the 3D
\ac{rcnn} model is able to effectively leverage the relationships between neighbouring voxels
within a patch. Notably, the relative performance of the 3D \ac{rcnn} decreases with
increasing subsampling ratios, compared to \ac{sh} interpolation. This indicates that within a
low subsampling regime, the learned joint kq-space distribution affords the 3D model additional
information, not present in the q-space distribution. In a higher sampling regime however, the
additional information present within the joint distribution is relatively diminished.
This implies that past a certain threshold the q-space distribution alone can be used to
effectively interpolate between points, without the need of additional spatial information
provided by the 3D model.

\begin{table}[H]
    \centering
    \resizebox{\columnwidth}{!}{
        \begin{tabular}{c||c|c||c|c||c|c||}
            & \multicolumn{2}{|c||}{$q_{in} = 6$, $q_{out} = 84$} & \multicolumn{2}{|c||}{$q_{in} = 10$, $q_{out} = 80$} & \multicolumn{2}{|c||}{$q_{in} = 30$, $q_{out} = 60$}\\ [0.5ex]
            \hline \hline
            Model & \ac{rmse} & \ac{mssim} & \ac{rmse} & \ac{mssim} & \ac{rmse} & \ac{mssim}\\
            \hline
            \ac{sh} Interpolation & $119.0\pm50.3$ & $0.9460\pm0.0419$ & $65.1\pm12.7$ & $0.9854\pm0.0044$ & $63.8\pm10.4$ & $0.9867\pm0.0033$\\
            1D \ac{rcnn} & $102.5\pm31.6$ & $0.9639\pm0.0225$ & $70.0\pm15.1$ & $0.9852\pm0.0054$ & $64.1\pm10.6$ & $0.9875\pm0.0033$\\
            3D \ac{rcnn} & \boldmath$78.4\pm13.9$ & \boldmath$0.9787\pm0.0071$ & \boldmath$63.4\pm12.5$ & \boldmath$0.9870\pm0.0040$ & \boldmath$63.4\pm10.2$ & \boldmath$0.9876\pm0.0032$\\
        \end{tabular}
    }
    \caption{
        Average performance of \ac{sar} in eight subjects with $b = 1000$ across different models. Best results
        are highlighted in bold.
    }
    \label{table:subsamp}
\end{table}

Table \ref{table:bvals} similarly shows that optimal performance is obtained from the individually
trained 3D \ac{rcnn} models, this time across all b-value shells. Additionally the combined 3D
model outperforms the 1D model across all shells and both metrics, whilst it has higher \ac{mssim}
across all shells and lower \ac{rmse} within the $b = 2000$ and $b = 3000$ shells compared to
\ac{sh} interpolation. In particular, \ac{sh} interpolation performance drops relative to all
three \ac{rcnn} models at b-values $b = 2000$ and $b = 3000$. Given that higher b-values yield
lower signal-to-noise ratios, this suggests that the deep learning models are potentially more
robust to noise, compared to the simpler \ac{sh} interpolation model. The effect of the shifted
distribution at higher b-values can be seen through the difference in \ac{rmse} and \ac{mssim}
values, independent of model. \ac{rmse}, which is not a normalised metric, decreases given an
increase in b-value, whereas this relationship is not present within the normalised \ac{mssim}
metric.

\begin{table}[H]
    \centering
    \resizebox{\columnwidth}{!}{
        \begin{tabular}{c||c|c||c|c||c|c||}
            & \multicolumn{2}{|c||}{$b = 1000$} & \multicolumn{2}{|c||}{$b = 2000$} & \multicolumn{2}{|c||}{$b = 3000$}\\ [0.5ex]
            \hline \hline
            Model & \ac{rmse} & \ac{mssim} & \ac{rmse} & \ac{mssim} & \ac{rmse} & \ac{mssim}\\
            \hline
            \ac{sh} Interpolation & $65.1\pm12.7$ & $0.9854\pm0.0044$ & $64.5\pm9.1$ & $0.9659\pm0.0088$ & $66.7\pm13.5$ & $0.9292\pm0.0242$\\
            1D \ac{rcnn} (Separate) & $70.0\pm15.1$ & $0.9852\pm0.0054$ & $51.3\pm6.7$ & $0.9766\pm0.0056$ & $48.1\pm8.0$ & $0.9566\pm0.0133$\\
            3D \ac{rcnn} (Separate) & \boldmath$63.4\pm12.5$ & \boldmath$0.9870\pm0.0040$ & \boldmath$48.1\pm6.2$ & \boldmath$0.9796\pm0.0045$ & \boldmath$42.6\pm5.5$ & \boldmath$0.9633\pm0.0088$\\
            3D \ac{rcnn} (Combined) & $65.7\pm13.2$ & $0.9869\pm0.0041$ & $49.0\pm6.0$ & $0.9788\pm0.0047$ & $44.8\pm6.2$ & $0.9616\pm0.0099$\\
        \end{tabular}
    }
    \caption{
        Average performance of \ac{sar} in eight subjects with $q_{in} = 10$ and $q_{out} = 80$. Best results
        are highlighted in bold.
    }
    \label{table:bvals}
\end{table}

\begin{figure}[!htb]
    \subfigure[Baseline]{
        \begin{overpic}[width=0.225\textwidth]{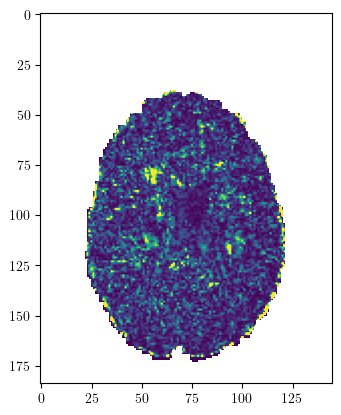}
            \put(13,90){\scriptsize WM: $0.0464$}
            \put(13,83){\scriptsize GM: $0.0501$}
        \end{overpic}
    }
    \hfill
    \subfigure[SH Interpolation]{
        \begin{overpic}[width=0.21\textwidth]{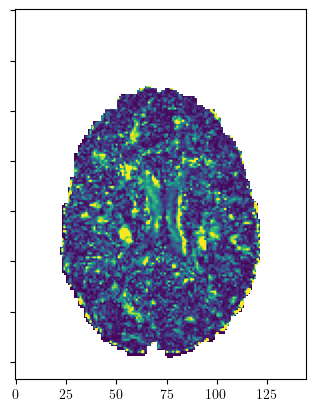}
            \put(7,90){\scriptsize WM: $0.0624$}
            \put(7,83){\scriptsize GM: $0.0549$}
        \end{overpic}
    }
    \hfill
    \subfigure[1D RCNN]{
        \begin{overpic}[width=0.21\textwidth]{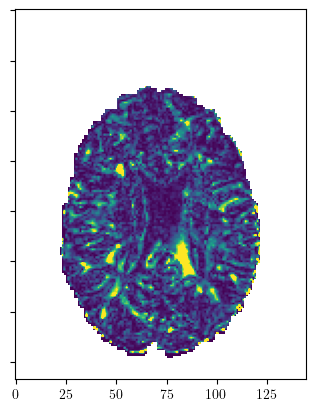}
            \put(7,91){\scriptsize WM: $0.0531$}
            \put(7,84){\scriptsize GM: $0.0503$}
        \end{overpic}
    }
    \hfill
    \subfigure[3D RCNN]{
        \begin{overpic}[width=0.261\textwidth]{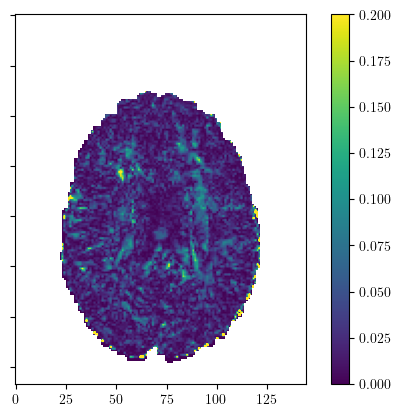}
            \put(7,90){\scriptsize WM: $0.0353$}
            \put(7,83){\scriptsize GM: $0.0357$}
        \end{overpic}
    }
    \caption{
        Axial slice of \ac{fa} \ac{ae} in one subject from the test dataset. \ac{sar} is
        performed with $q_{in} = 6$, $q_{out} = 84$. \ac{wm} and \ac{gm} values are averaged
        across voxels only within the \ac{wm} and \ac{gm} mask, respectively. The baseline
        \ac{fa} map is calculated from $q_{in}$ volumes whilst other \ac{fa} maps are derived from
        both $q_{in}$ and $q_{out}$ data.
    }
    \label{fig:fa_compare}
\end{figure}
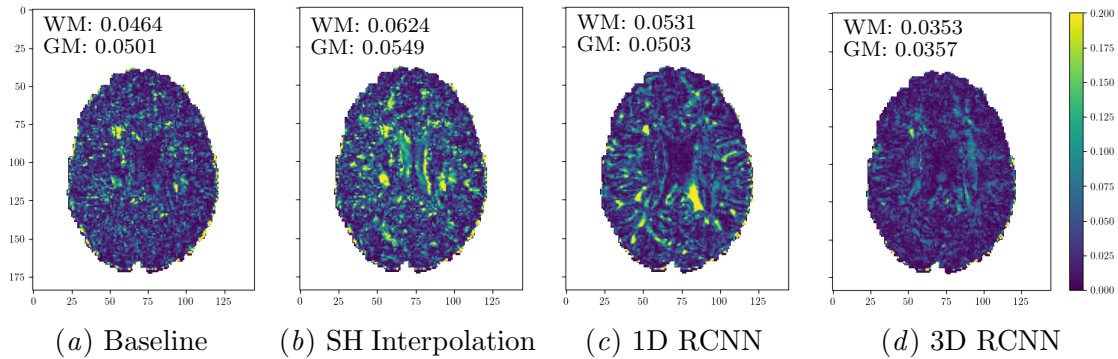

Figure \ref{fig:fa_compare} shows an axial slice of \ac{fa} \ac{ae} within one subject across
different models. Here \ac{ae} is visibly lowest within the 3D \ac{rcnn} model, whilst the baseline
derived \ac{ae} is lower than in the \ac{sh} and 1D models. This trend is also true when segmenting
the axial slice into \ac{wm} \& \ac{gm} voxels and analysing the performance separately in each
tissue type. The relatively low baseline error is likely due to \ac{dti} not requiring
\ac{hardi}, and therefore is robust even at low q-space sampling rates. Figure
\ref{fig:model_compare} presents the \ac{rmse} of the inferred \ac{dmri} data used to derive the
\ac{fa} maps, and are consistent with the findings in Figure \ref{fig:fa_compare}. In particular,
the lower relative \ac{wm} error rates present within the 3D model is important for downstream
analysis techniques that require \ac{hardi} as they often focus on voxels containing a high
proportion of \ac{wm}. A breakdown of \ac{wm} and \ac{gm} \ac{rmse} across different subsampling
ratios and b-values can be found in appendix \ref{sec:ts_rmse}, which contain similar trends as
those presented in Tables \ref{table:subsamp} and \ref{table:bvals}. \ac{wm} and \ac{gm} masks are
generated using FSL FAST \citep{zhang2000segmentation}, whilst \ac{dti} metrics are generated using
the FSL Diffusion Toolbox.

\section{Conclusion and Future Work}

We present a recurrent 3D convolutional architecture to perform angular super-resolution on
diffusion MRI data. We compare this methodology against a relevant angular interpolation technique,
as well as a 1D variant of the architecture. We demonstrate that the 3D model performs best across
various subsampling ratios and b-values. Additionally, we show that this architecture can be used
to train a model capable of inferring several different b-values concurrently, albeit at slightly
reduced performance compared to individually trained models.

Further work is needed to quantify the robustness of this methodology in out-of-distribution
datasets such as those with pathologies and different acquisition parameters, and to
provide a comparison with non-recurrent convolutional architectures. A future extension
to this work would be expanding the model to explicitly infer other shells, thereby
performing multi-shell angular super-resolution. When expanding this model, the effect of the
subsampling ratio on multi-shell inference should be explored. Additionally, future work should
investigate the effect of angular super-resolution on downstream single-shell and multi-shell
analyses that require high angular resolutions.

\midlacknowledgments{
    This work was conducted at the Univeristy of Sheffield whilst authors Matthew Lyon and
    Mauricio A Álvarez were affiliated with the organisation. This work was funded by the
    Engineering and Physical Sciences Research Council (EPSRC) Doctoral Training Partnership
    (DTP) Scholarship. Data were provided [in part] by the \ac{hcp}, WU-Minn Consortium
    (Principal Investigators: David Van Essen and Kamil Ugurbil; 1U54MH091657) funded by the 16
    NIH Institutes and Centers that support the NIH Blueprint for Neuroscience Research; and by the
    McDonnell Center for Systems Neuroscience at Washington University.
}

\bibliography{lyon22}

\appendix
\counterwithin{figure}{section}
\counterwithin{table}{section}

\section{Model Architecture}

\begin{center}
    \includegraphics[width=\textwidth]{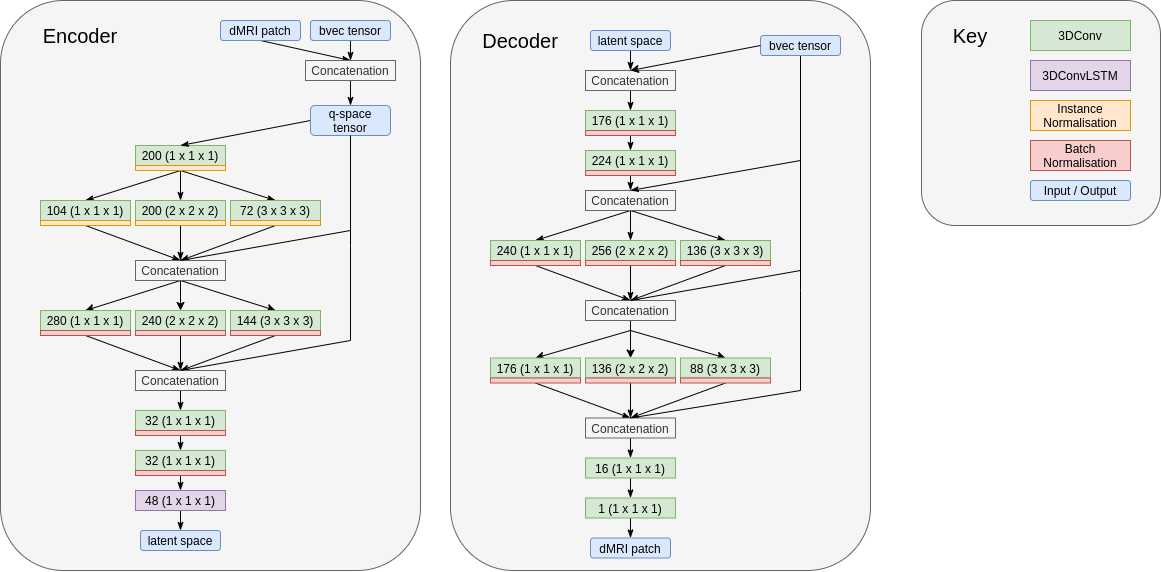}
    \captionof{figure}{
        \ac{rcnn} model diagram with convolutional filter sizes and channel dimensions. Each
        convolution node specifies the number of filters used (left) and filter size (right).
    }
    \label{fig:rcnn_full}
\end{center}

\section{Tissue Specific RMSE} \label{sec:ts_rmse}

\begin{figure}[!htb]
    \subfigure[SH Interpolation]{
        \begin{overpic}[width=0.3\textwidth]{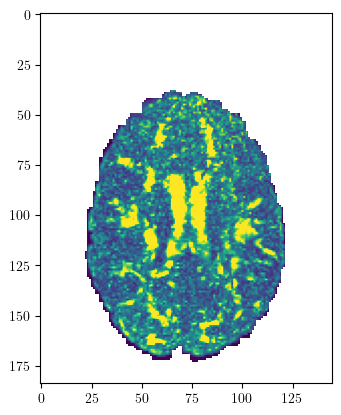}
            \put(12,90){\scriptsize WM: $180.0$}
            \put(12,83){\scriptsize GM: $124.6$}
        \end{overpic}
    }
    \hfill
    \subfigure[1D RCNN]{
        \begin{overpic}[width=0.279\textwidth]{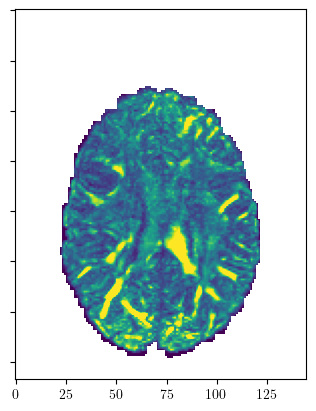}
            \put(6,91){\scriptsize WM: $140.0$}
            \put(6,84){\scriptsize GM: $113.9$}
        \end{overpic}
    }
    \hfill
    \subfigure[3D RCNN]{
        \begin{overpic}[width=0.342\textwidth]{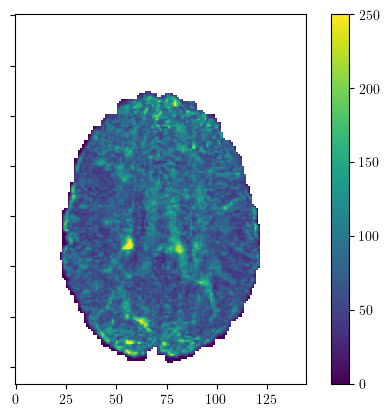}
            \put(6,90){\scriptsize WM: $92.2$}
            \put(6,83){\scriptsize GM: $84.5$}
        \end{overpic}
    }
    \caption{
        Axial slice of \ac{rmse} in one subject from the test dataset. \ac{sar} is performed with
        $q_{in} = 6$, $q_{out} = 84$. Here each \ac{rmse} pixel is the \ac{rmse} averaged across
        $q_{out}$ directions. \ac{wm} and \ac{gm} values are the \ac{rmse} averaged across spatial
        voxels and $q_{out}$ directions, for voxels only within the \ac{wm} and \ac{gm}
        mask, respectively.
    }
    \label{fig:model_compare}
\end{figure}
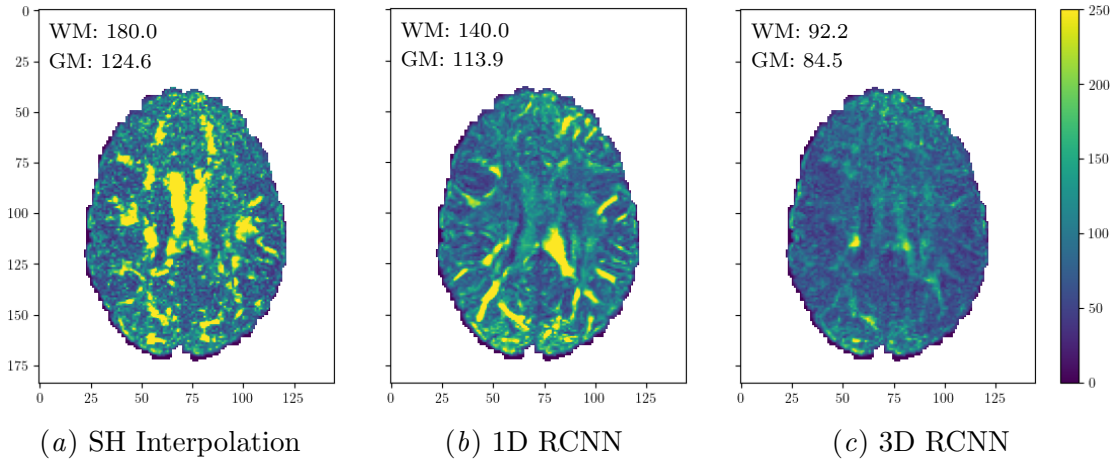

\begin{table}[H]
    \centering
    \resizebox{\columnwidth}{!}{
        \begin{tabular}{c||c|c||c|c||c|c||}
            & \multicolumn{2}{|c||}{$q_{in} = 6$, $q_{out} = 84$} & \multicolumn{2}{|c||}{$q_{in} = 10$, $q_{out} = 80$} & \multicolumn{2}{|c||}{$q_{in} = 30$, $q_{out} = 60$}\\ [0.5ex]
            \hline \hline
            Model & \ac{wm} & \ac{gm} & \ac{wm} & \ac{gm} & \ac{wm} & \ac{gm}\\
            \hline
            \ac{sh} Interpolation & $142.8\pm67.9$ & $109.5\pm42.8$ & $67.3\pm9.7$ & \boldmath$65.9\pm15.2$ & $63.8\pm6.9$ & \boldmath$65.7\pm12.5$\\
            1D \ac{rcnn} & $116.5\pm41.5$ & $100.4\pm29.2$ & $68.1\pm13.2$ & $73.8\pm17.1$ & $60.3\pm7.9$ & $67.9\pm12.3$\\
            3D \ac{rcnn} & \boldmath$82.9\pm15.4$ & \boldmath$79.1\pm14.5$ & \boldmath$61.0\pm9.1$ & $66.5\pm14.4$ & \boldmath$59.1\pm7.2$ & $67.3\pm11.8$\\
        \end{tabular}
    }
    \caption{
        Average \ac{wm} and \ac{gm} \ac{rmse} in eight subjects with $b = 1000$ across different models. Best
        results are highlighted in bold.
    }
    \label{table:subsamp_wmgm}
\end{table}

\begin{table}[H]
    \centering
    \resizebox{\columnwidth}{!}{
        \begin{tabular}{c||c|c||c|c||c|c||}
            & \multicolumn{2}{|c||}{$b = 1000$} & \multicolumn{2}{|c||}{$b = 2000$} & \multicolumn{2}{|c||}{$b = 3000$}\\ [0.5ex]
            \hline \hline
            Model & \ac{wm} & \ac{gm} & \ac{wm} & \ac{gm} & \ac{wm} & \ac{gm}\\
            \hline
            \ac{sh} Interpolation & $67.3\pm9.7$ & \boldmath$65.9\pm15.2$ & $81.1\pm12.7$ & $57.1\pm8.1$ & $88.0\pm19.4$ & $56.1\pm10.4$\\
            1D \ac{rcnn} (Separate) & $68.1\pm13.2$ & $73.8\pm17.1$ & $56.2\pm6.9$ & $50.6\pm7.4$ & $56.7\pm10.6$ & $45.6\pm7.1$\\
            3D \ac{rcnn} (Separate) & \boldmath$61.0\pm9.1$ & $66.5\pm14.4$ & \boldmath$51.6\pm5.7$ & \boldmath$48.2\pm7.0$ & \boldmath$48.5\pm7.2$ & \boldmath$41.2\pm5.1$\\
            3D \ac{rcnn} (Combined) & $62.8\pm9.8$ & $69.5\pm15.2$ & $52.9\pm5.8$ & $48.8\pm6.8$ & $52.1\pm8.2$ & $42.5\pm5.5$\\
        \end{tabular}
    }
    \caption{
        Average \ac{wm} and \ac{gm} \ac{rmse} in eight subjects with $q_{in} = 10$ and $q_{out} = 80$. Best
        results are highlighted in bold.
    }
    \label{table:bvals_wmgm}
\end{table}

\section{Comparison to Non-Recurrent Architecture}

We compare performance of the 3D \ac{rcnn} architecture against a similar 3D \ac{cnn}
design. Here the \ac{cnn} models use a 1D convolutional layer in place of the
\ac{clstm} layer, whilst maintaining all other architecture hyperparameters. Table
\ref{table:subsamp_cnn} compares performance across q-space subsampling ratios, whilst Table
\ref{table:bvals_cnn} compares across b-values. The \ac{rcnn} performs best across all metrics,
suggesting that the added internal complexity of the recurrent layer provides additional capacity
for the architecture to capture the non-trivial relationship of q-space in the data.

\begin{table}[H]
    \centering
    \resizebox{\columnwidth}{!}{
        \begin{tabular}{c||c|c||c|c||c|c||}
            & \multicolumn{2}{|c||}{$q_{in} = 6$, $q_{out} = 84$} & \multicolumn{2}{|c||}{$q_{in} = 10$, $q_{out} = 80$} & \multicolumn{2}{|c||}{$q_{in} = 30$, $q_{out} = 60$}\\ [0.5ex]
            \hline \hline
            Model & \ac{rmse} & \ac{mssim} & \ac{rmse} & \ac{mssim} & \ac{rmse} & \ac{mssim}\\
            \hline
            3D \ac{cnn} & $84.8\pm15.9$ & $0.9758\pm0.0085$ & $68.3\pm12.6$ & $0.9855\pm0.0043$ & $66.4\pm10.4$ & $0.9873\pm0.0032$\\
            3D \ac{rcnn} & \boldmath$78.4\pm13.9$ & \boldmath$0.9787\pm0.0071$ & \boldmath$63.4\pm12.5$ & \boldmath$0.9870\pm0.0040$ & \boldmath$63.4\pm10.2$ & \boldmath$0.9876\pm0.0032$\\
        \end{tabular}
    }
    \caption{
        Average performance of \ac{sar} in eight subjects with $b = 1000$ across \ac{cnn} and \ac{rcnn} models.
        Best results are highlighted in bold.
    }
    \label{table:subsamp_cnn}
\end{table}

\begin{table}[H]
    \centering
    \resizebox{\columnwidth}{!}{
        \begin{tabular}{c||c|c||c|c||c|c||}
            & \multicolumn{2}{|c||}{$b = 1000$} & \multicolumn{2}{|c||}{$b = 2000$} & \multicolumn{2}{|c||}{$b = 3000$}\\ [0.5ex]
            \hline \hline
            Model & \ac{rmse} & \ac{mssim} & \ac{rmse} & \ac{mssim} & \ac{rmse} & \ac{mssim}\\
            \hline
            3D \ac{cnn} & $68.3\pm12.6$ & $0.9855\pm0.0043$ & $50.0\pm6.3$ & $0.9779\pm0.0048$ & $70.7\pm12.7$ & $0.9350\pm0.0157$\\
            3D \ac{rcnn} & \boldmath$63.4\pm12.5$ & \boldmath$0.9870\pm0.0040$ & \boldmath$48.1\pm6.2$ & \boldmath$0.9796\pm0.0045$ & \boldmath$42.6\pm5.5$ & \boldmath$0.9633\pm0.0088$\\
        \end{tabular}
    }
    \caption{
        Average performance of \ac{sar} in eight subjects with $q_{in} = 10$ and $q_{out} = 80$ in \ac{cnn} and
        \ac{rcnn} models. Best results are highlighted in bold.
    }
    \label{table:bvals_cnn}
\end{table}

\end{document}